\documentclass{article}
\usepackage{arxiv}
\usepackage{amsmath}
\usepackage{cite}
\usepackage[utf8]{inputenc} 
\usepackage[T1]{fontenc}    
\usepackage{url}            
\usepackage{booktabs}       
\usepackage{amsfonts}       
\usepackage{nicefrac}       
\usepackage{microtype}      
\usepackage{lipsum}
\usepackage{graphicx}
\usepackage{dcolumn}
\usepackage{bm}
\usepackage{stackengine}
\usepackage{xcolor}
\usepackage{scalerel}
\usepackage{tikz}
\usetikzlibrary{svg.path}
\definecolor{orcidlogocol}{HTML}{A6CE39}
\tikzset{
  orcidlogo/.pic={
    \fill[orcidlogocol] svg{M256,128c0,70.7-57.3,128-128,128C57.3,256,0,198.7,0,128C0,57.3,57.3,0,128,0C198.7,0,256,57.3,256,128z};
    \fill[white] svg{M86.3,186.2H70.9V79.1h15.4v48.4V186.2z}
                 svg{M108.9,79.1h41.6c39.6,0,57,28.3,57,53.6c0,27.5-21.5,53.6-56.8,53.6h-41.8V79.1z M124.3,172.4h24.5c34.9,0,42.9-26.5,42.9-39.7c0-21.5-13.7-39.7-43.7-39.7h-23.7V172.4z}
                 svg{M88.7,56.8c0,5.5-4.5,10.1-10.1,10.1c-5.6,0-10.1-4.6-10.1-10.1c0-5.6,4.5-10.1,10.1-10.1C84.2,46.7,88.7,51.3,88.7,56.8z};
  }
}

\newcommand\orcidicon[1]{\href{https://orcid.org/#1}{\mbox{\scalerel*{
\begin{tikzpicture}[yscale=-1,transform shape]
\pic{orcidlogo};
\end{tikzpicture}
}{|}}}}
\usepackage{hyperref} 

\title{Generalized entropies, density of states, and non-extensivity}

\author{
 S\'{a}muel G. Balogh \orcidicon{0000-0003-3417-2655}\\
 Dept. of Biological Physics, E\"{o}tv\"{o}s University,\\
 H-1117 Budapest, Hungary \\
 \texttt{balogh@hal.elte.hu} \\
   \AND
  Gergely Palla \orcidicon{0000-0002-3406-4200}\\
  MTA-ELTE Statistical and Biological Physics Research Group,\\ Dept. of Biological Physics, E\"{o}tv\"{o}s University,\\ H-1117 Budapest, Hungary
  \And
  P\'{e}ter Pollner \orcidicon{0000-0003-0464-4893}\\
  MTA-ELTE Statistical and Biological Physics Research Group,\\ Dept. of Biological Physics, E\"{o}tv\"{o}s University,\\ H-1117 Budapest, Hungary
  \And
  D\'{a}niel Cz\'{e}gel \orcidicon{0000-0002-5722-1598}\\
Institute of Evolution, Centre for Ecological Research, H-8237 Tihany, Hungary\\
Dept. of Plant Systematics, Ecology and Theoretical Biology, E\"{o}tv\"{o}s University, H-1117 Budapest, Hungary\\
Center for the Conceptual Foundations of Science, Parmenides Foundation, 82049 Pullach/Munich, Germany\\
 \texttt{danielczegel@gmail.com} \\
}

\begin{document}
\maketitle
\begin{abstract}

The concept of entropy connects the number of possible configurations with the number of variables in large stochastic systems. Independent or weakly interacting variables render the number of configurations scale exponentially with the number of variables, making the Boltzmann-Gibbs-Shannon entropy extensive. In systems with strongly interacting variables, or with variables driven by history-dependent dynamics, this is no longer true. 
Here we show that contrary to the generally held belief, not only strong correlations or history-dependence, but skewed-enough distribution of visiting probabilities, that is, first-order statistics, also play a role in determining the relation between configuration space size and system size, or, equivalently, the extensive form of generalized entropy. We present a macroscopic formalism describing this interplay between first-order statistics, higher-order statistics, and configuration space growth. We demonstrate that knowing any two strongly restricts the possibilities of the third.
We believe that this unified macroscopic picture of emergent degrees of freedom constraining mechanisms provides a step towards finding order in the zoo of strongly interacting complex systems.


 
\end{abstract}


\section{Introduction}

Today, witnessing the feedback loop of developing digital technologies and increasing amount of data collected, there has been an ever increasing need and opportunity to understand and control complex biological, social or technological systems \cite{HT_complex_sytems,Tsallis_book,Tsallis_complex_systems,Tsallis_open_quest,beyondbigdata}. The hallmark of such systems is that their global behavior emerges out of a large number of stochastic variables interacting in a non-trivial way \cite{Corominas_murtra_history_dep,correlated_thermo,micromacro,ruseckas_binary_spin}. A useful level of description is provided by (generalized) statistical mechanics, an effort to identify relations between relevant observable summary statistics of stochastic dynamics over configuration space. In many cases, the microscopic dynamical rules governing the system are not known; instead, their effect on first-order and higher-order statistics (i.e., visiting probabilities and spatial/temporal correlations) over configuration space form the basis of understanding. A first classification of all systems is given by the mere \emph{size} of the configuration space $W$ in the function of the number of microscopic variables $N$. This is also a necessary classification for system-size independent modeling. In the absence of interactions, $W(N)$ grows exponentially as the joint distribution over microscopic variables factorize. Non-trivial joint distributions, however, result in non-trivial restrictions on configuration space and possibly non-exponential scaling of $W(N)$. Such systems are labelled non-extensive. In this paper, we factor sources of non-extensivity to first-order and higher-order statistical properties of the joint distribution over microscopic variables. In particular, effective classification of all higher-order statistics, however complicated they are, have been introduced under the name of \emph{generalized entropies}. Generalized entropies indirectly model correlations: the specific entropic form that scales with system size (i.e., extensive, $S\sim N$), tells us which class the system itself belongs to. 

A variety of generalized entropic functionals have been introduced to phenomenologically extend statistical mechanics to specific non-ergodic or strongly interacting systems, both within and outside the realm of physics including spin-like systems \cite{ruseckas_binary_spin,ruseckas_binary_spin_2,Tempesta_super_exp}, cosmic ray energy spectra \cite{beck_scirep}, multifractals \cite{Gadjiev_2015}, networks \cite{Tsallis_geographical}, quantum information \cite{ch_scirep,quantum_info,Shafee2}, special relativity \cite{Kaniadakis}, anomalous diffusive processes \cite{Chavanis_anomalous_diff,Plastino_fokker_planck,HT_aging_random_walks,czegel_phase_space_volume_scaling,curado_two_temperature}, superstatistics \cite{tsallis_gen_ent_superstat,Hanel_superstat}, time series analysis \cite{entropy_application_time_series1,entropy_application_time_series2,amigo_time_series} and artificial neural networks \cite{neural_networks}. 

The diversity of proposed entropic functionals reflects the conceptual diversity behind the assumptions all leading, in weakly correlated systems, to the same mathematical form of the Boltmann-Gibbs-Shannon entropy, $S_{\text{BGS}}=\sum_{i=1}^W - p_i \ln p_i$. In particular, arguments relying on thermodynamics, statistical mechanics, dynamical systems, information theory, and statistics all provide means to derive $S_{\text{BGS}}$ as a useful measure, and they all provide different means to generalize it \cite{Tsallis_book,HT_gen_ent_class,tempesta_scirep,Tempesta_gen_ent,Kaniadakis,Shafee,Bizet_Splusminus,Korbel_gen_ent_class,h_derivative_ent,hypoentropy,gen_entropy_review,Tempesta_group_ent}. Here we do not commit to any of these conceptual frameworks, instead, following the work by Hanel and Thurner in Ref.\cite{HT_gen_ent_class}, we rely on an axiomatic characterization of generalized entropies, based on the Shannon-Khinchin axioms SK1-SK4 \cite{khinchin}. Assuming that most of the relevant generalized entropic forms can be written as a sum of a \emph{pointwise} function $g$ over probabilities,
\begin{equation}
    S_{g}\left[p\right]=\sum\limits^{W}_{i=1}g\left(p_i\right)
\end{equation}
with a notable exception being the class of R\'{e}nyi entropies, $S_{\text{R\'{e}nyi}}=\frac{1}{1-\alpha}\log\sum_{i=1}^W p_i^{\alpha}$, axioms SK1-SK4 regarding $S_g[p]$ translate to the language of the entropic kernel $g(p)$. Prescribing all SK1-SK4 uniquely determines $g$ to be proportional to the Boltzmann-Gibbs-Shannon kernel, $g_{\text{BGS}}=-p \ln p$. A surprisingly rich phenomenology of all possible generalizations to non-extensive systems can be achieved by discarding the only SK axiom that prescribe the resulting entropy to be additive, namely, the decomposability axiom SK4, $S[p_{AB}]=\langle S[p_{A|B}] \rangle_B + S[p_B]$, expressing that the entropy of a joint distribution $p_{AB}$ can be decomposed to the expected entropy of the conditional distribution $p_{A|B}$ and the entropy of the marginal $p_B$.

Interestingly, assuming equiprobable configurations ($p_i\equiv W^{-1}$), all possible entropies obeying to SK1-SK3 follow the asymptotic scaling law 
\begin{equation}
   R_{\lambda}=\lim_{W\to\infty}\frac{S_g[p_i\equiv\frac{1}{\lambda W}]}{S_g[p_i\equiv\frac{1}{W}]}\sim \lambda^{1-c}
   \label{eq:Hannel_Turner_scaling_base_def}
\end{equation}
with Hanel-Thurner (H-T) exponent $0 < c \leq 1$ \cite{HT_gen_ent_class}. A particularly simple representative of \emph{each} possible asymptotic non-extensivity class is given by the one-parameter family of Tsallis entropies \cite{Tsallis_first}, $g_{q}(p)=\frac{p-p^q}{q-1}$, with $g_q$ belonging to the class $c=q$, limiting $g_{\text{BGS}}$ (class $c=1$) as $q\to 1$.

Crucially, for any specific system, the functional form of the generalized entropy that is extensive maps \emph{any} correlation structure beyond first-order statistics $\{p_i\}$ to its global consequences: the scaling of configuration space $W(N)$ with  system size, classified by the H-T exponent $c$. In this paper, we unify this phenomenological description with the effect of first-order statistics to gain a complete picture of sources of non-extensive configuration space growth, factored to first-order and higher-order statistics. In particular, as entropies are invariant to relabeling of states $i\leftrightarrow j$, they only depend on the fraction of states with probability $p$. This probability density over probabilities we call \emph{density of states} $\varrho(p)$, similarly to statistical physics and condensed matter theory, where densities over log-probabilities play an important role. 

The paper is organized as follows. The Results section introduces the general formalism, subsection Density of states: examples discusses the role of specific density of states, and the Discussion summarizes the results. Detailed calculations are given in the Supplementary Information S2.

The paper is organized as follows. The Results section introduces the general formalism, subsection Density of states: examples discusses the role of specific density of states, and the Discussion summarizes the results. Detailed calculations are given in the Supplementary Information S2.

\section*{Results}\label{sect:results}

\begin{figure}[htp]
  \begin{center}
  \centering
    \includegraphics[width=1.0\textwidth]{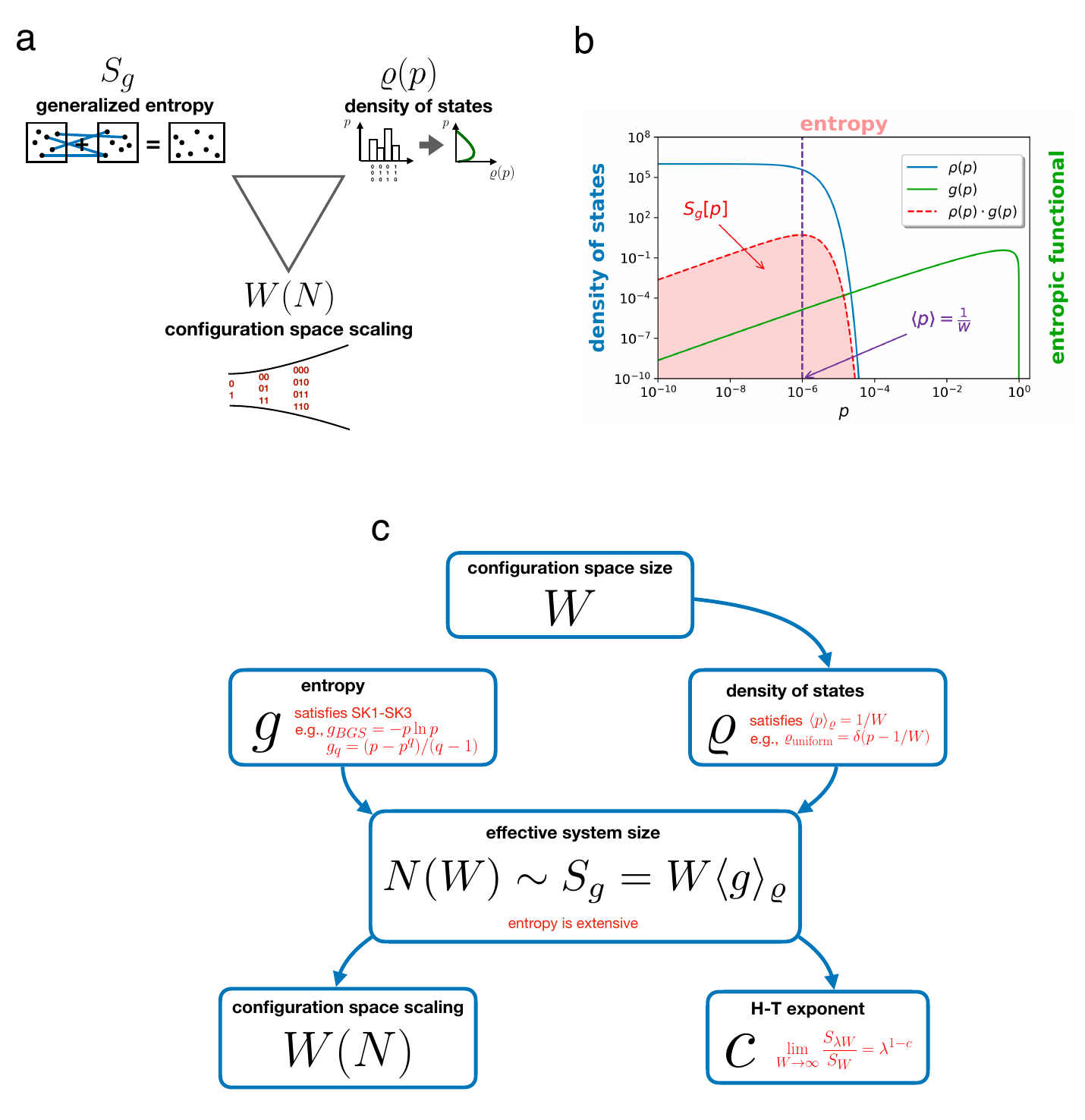}
      \caption{\textbf{a)} Generalized entropies $S_g$, providing a phenomenological classification of higher order statistics over configuration space, density of states $\varrho(p)$, summarizing first order statistics over configuration space , and configuration space scaling with system size $W(N)$. Knowledge of any two strongly restricts the possibilities for the third. \textbf{b)} Computation of entropy $S_g$, corresponding to the shaded area, based on kernel $g$ and density of states $\varrho$. In this example, $g=g_{\text{BGS}}=-p\ln p$, $\varrho(p)$ is exponential, and the size of the configuration space is $W=10^6$, corresponding to $\left< p\right>_{\varrho} =W^{-1}=10^{-6}$. \textbf{c)} Computational steps relating generalized entropy $S_g$, density of states $\varrho$, configuration space scaling $W(N)$, and Hanel-Thurner exponent $c$. Density of states and generalized entropies summarize first and higher order statistics over configuration space, respectively, whereas configuration space scaling and the H-T exponent classify complex systems based on how available configuration space scales with effective system size $N$. Note that the starting point is the size of configuration space $W$; effective system size $N$ is determined by leveraging extensivity of the system-specific generalized entropic form.}
    \label{fig1}
    \end{center}
\end{figure}

In this section, we develop a simple mathematical framework relating three central concepts of strongly correlated systems: first order statistics, quantified in terms of density of states $\varrho(p)$, higher-order statistics, modeled by generalized entropic functionals $S_g[p]$ and the scaling of configuration space with system size $W(N)$, classified by the H-T exponent $c$. Figure \ref{fig1}a illustrates the idea.

While doing so, we attempt to provide a step-by-step introduction to the logic of generalized statistical mechanics. Generalized statistical mechanics phenomenologically classifies \emph{all} conceivable correlated systems by compressing all statistical dependences in the system's stochastic dynamics into one relevant measure: the scaling of configuration space with system size. It uses a reverse logic: starting from the size of the configuration space $W$, through the prescription of extensivity of system-specific generalized entropic functional, one arrives to the  system size $N(W)$ \cite{Scarfone_extensivity,HT_gen_ent_appl}. Note that based on observing (some statistics of) the dynamics over configuration space, this is a meaningful definition of system size, whereas the mere number of variables is not: think of a configuration space defined by many copies of the same variable (i.e., maximal mutual information between them). In order to avoid confusion, from now on we refer to $N$ as \emph{effective system size}. Inverting (the asymptotics of) $N(W)$ tells us the system's configuration space scaling $W(N)$, classified by the H-T exponent $c$ through Eq. (\ref{eq:Hannel_Turner_scaling_base_def}). We complement this algorithmic recipe, depicted in Figure \ref{fig1}c, with the missing ingredient, first-order statistics $\varrho(p)$, to yield a coherent picture of all possible sources of non-extensivity, factored to first and higher-order correlations.

Asymptotically, $W\to\infty$, any generalized entropy $S_g$ can be written as
\begin{equation}
    S_{g}\left[p\right]=\sum\limits^{W}_{i=1}g\left(p_i\right)=W \int\limits_0^1 \varrho(p) g(p) \mathrm{d}p=W \left < g(p) \right>_{\varrho}
\label{eq:rho_gen_entropies}
\end{equation}
by grouping terms with the same probability in the sum, weighted by the density of states $\varrho$, visualized in Figure \ref{fig1}b. Concavity of $g$, along with Jensen's inequality, $\left< g\left(p\right)\right >_{\varrho} \leq g\left(\left< p \right>_{\varrho}\right)$, guarantees that $S_g$ is maximal for the uniform distribution over states, $\varrho(p)=\delta(p-1/W)$ (see Supplementary Information S2).

The density of states $\varrho$ cannot be arbitrary, however: it is constrained by the normalization condition on $p$,
\begin{equation}
    1=\sum_i p_i=W\int\limits_0^1 \varrho(p) p \mathrm{d}p=W \left< p \right >_{\varrho},
\label{eq:rho_exp_val_const}
\end{equation}
that is, the expected value of $p$ under $\varrho$ is fixed to be $1/W$, decreasing the dimension of the parameter space of $\varrho$ by $1$. When needed, we emphasize this constraint by explicitly writing $\varrho(p|W)$. An additional technical difficulty stems from the fact that the support of $\varrho$ is bounded. In this paper, we choose density of states that are bounded on $[0,1]$ but limit well-known distributions over a semi-infinite support as $W\to\infty$, and consequently, as $\left< p \right >_{\varrho}\to 0$.

The particular forms of $\varrho$ we consider, with detailed calculations in subsection of Density of states: examples and in the Supplementary Information S2, are i) a delta function $\varrho(p)=\delta(p-1/W)$, corresponding to uniform distribution over states (\emph{microcanonical, MC}), ii) a combination of multiple delta functions, describing multiple uniform domains in configuration space, possibly scaling differently with configuration space size $W$ (\emph{multi-delta, MD}), iii) a case in which a single state has macroscopic (non-disappearing) probability at the limit $W\to\infty$ and the probability of all other states are equal (\emph{Bose-Einstein, BE}), iv) an exponential density of states (\emph{exponential}), v) a log-gamma density of states that limits log-normal (\emph{log-gamma}), one that is a power-law with exponent limiting $-1$ (\emph{power law}), and vi) a two-parameter family over $[0,1]$, the beta distribution, where we tune the one remaining free parameter to achieve a power-law tail with a tunable exponent (\emph{beta}).

\subsection*{Configuration space scaling}
\label{sect:sample_space_scaling}

A simple classification of all correlated systems can be given by assessing how the associated extensive generalized entropy ($S_g\sim N$) reacts to configuration space rescaling, $W\to\lambda W$. This idea, consistent with the Shannon-Khinchin axiomatic foundations of information, is articulated in terms of the Hanel-Thurner (H-T) exponent $c$, defined in Eq. (\ref{eq:Hannel_Turner_scaling_base_def}). Here we generalize H-T scaling to systems with arbitrary (not necessarily uniform) visiting probabilities over states as
\begin{align}
     R_{\lambda}=\lim_{W\to\infty}\frac{S_g\left[\{p_i\}|\left< p\right>_{\varrho}=\frac{1}{\lambda W}\right]}{S_g\left[\{p_i\}|\left< p\right>_{\varrho}=\frac{1}{W}\right]}=\frac{\lambda W\int\limits_{0}^{1}\varrho(q|\lambda W)g(q)\mathrm{d}q}{ W\int\limits_{0}^{1}\varrho(q| W)g(q)\mathrm{d}q}\sim\lambda^{1-c}.
\label{eq:gen_scaling}
\end{align}
H-T scaling in case of a uniform distribution over configuration space is recovered by setting $\varrho(p|W)=\delta(p-1/W)$, simplifying Eq. (\ref{eq:gen_scaling}) to 
    $R_{\lambda}=\lim_{W\to\infty}\lambda g\left(\frac{1}{\lambda W}\right) / g\left(\frac{1}{W}\right)\sim \lambda^{1-c},
    \label{eq:HTscaling2}$ 
implying that $g(p)$ scales as $p^c$ as $p\to 0^+$. For example, $g_{\text{BGS}}=-p\ln p\sim p$, and $g_q=\frac{p-p^q}{q-1}\sim p^q$ if $0<q\leq 1$, in agreement with $S_{\text{BGS}}$ and $S_q$ belonging to H-T class $c=1$ and $c=q$, respectively. In a general framework, however, non-extensivity, classified by the H-T exponent $c$, depends on both first order and higher order statistics, accounted for by $\varrho$ and $g$ jointly. In particular, as the density of states $\varrho$ broadens (while still obeying to the expected value constraint $\left< p\right>_{\varrho} =W^{-1}$), contributions to the total entropy come from configurations with a wider range of probabilities. Figure \ref{fig2}b shows the contribution of configurations with probability $p<r$ to the total entropy, which we call \emph{cumulative entropy} $\Pi_g(r)$,

\begin{equation}
\Pi_g(r)=\frac{\int\limits^{r}_0\varrho(p)g(p)\mathrm{d}p}{\int\limits^{1}_0\varrho(p)g(p)\mathrm{d}p},
\label{eq:cum-entropy}
\end{equation}
for specific combinations of density of states $\varrho$ and entropy kernels $g$.

How does broadening of $\varrho$ affect the \emph{scaling} of available configuration space $W(N)$, and its classification, given by the H-T exponent $c$? In the following, we perform exact calculations following the steps illustrated on Figure \ref{fig1}c, for specific density of states $\varrho$ and entropy kernels $g$. We summarize the results in Table\ref{tab:bgs_tab} for BGS entropy and in Table\ref{tab:tsallis_tab} for Tsallis entropies.

\subsection*{Density of states: examples}\label{subsect:dos}
 
In order to keep track of the consequences of changing the density of states $\varrho$ alone, we introduce the following nomenclature. We refer to the scaling of $S_{g}=W\left < g(p)\right >_{\varrho}$ with $W$ as \emph{regular} if
\begin{equation}
W\left < g(p)\right >_{\varrho} \stackrel{\ W\to\infty \ }{\simeq} C_0+C_1W\left < g(p)\right >_{\delta}
\label{eq:regularanomalous}
\end{equation}
with $C_1\neq 0$, where $\delta$ denotes uniform configuration probabilities over the sample space (microcanonical ensemble). Otherwise, the scaling is referred to as $\emph{anomalous}$. 

\subsubsection*{Uniform distribution over configuration space (microcanonical)}
The simplest example is associated with the microcanonical ensemble whose density of states is written as $\varrho(p)=\delta(p-1/W)$. This form enables us to analitically express all forms of generalized entropies as
\begin{equation}
    S_g=W\int\limits_0^1 \delta\left(p-\frac{1}{W}\right) g(p) \mathrm{d}p=Wg\left(\frac{1}{W}\right)
    \label{eq:micro_entropy}.
\end{equation}
Based on that, e.g. the $S_{\text{BGS}}\sim \ln W$ and $S_q\sim\frac{1-W^{1-q}}{q-1}$ dependence for the Boltzmann-Gibbs-Shannon and Tsallis entropies can be shown in a straightforward manner. Hence, extensivity of $S_{\text{BGS}}$ is ensured by imposing exponential configuration space scaling $W(N)\sim e^N$, whereas $S_q$ is extensive under $W(N)\sim N^{\frac{1}{1-q}}$ (for further details see Table\ref{tab:bgs_tab} and Table\ref{tab:tsallis_tab}). 

As the upper bound of $S_q$ is always realized by the microcanonical ensemble, this case is corresponding to maximal disorder, which coincides with the SK2 \emph{maximality} axiom.

\subsubsection*{Multiple uniform domains over configuration space (multi-delta)}

A straightforward generalization of the classical microcanonical ensemble, 
let us discuss a system whose phase space is decomposeable into several disjunct $k+1$ sub-domains (e.g., $k+1$ different set of configurations). The volume of these sub-domains might scale differently with the size of the system being denoted by $V_0(N),V_1(N),...,V_k(N)$. We additionally assume that the $V_0(N)$ function is standing for the scaling of a forbidden region, and each configuration corresponding to the same $n$-th set of configurations is occurring with a probability proportional to the size of this population $\sim V_n$. Hence, the DOS with the correct pre-factors can simply be formulated as
\begin{equation}
    \varrho(p)=\frac{V_0}{W}\delta(p)+\sum\limits^k_{j=1} \frac{V_j}{W}\delta\left(p-\frac{V_j}{\sum^k_{m=1}V^2_m}\right),
    \label{eq:multidelta}
\end{equation}
where $\sum^k_{j=0}V_j(N)=W(N)$. In the most trivial case of $k=1$, Eq. (\ref{eq:multidelta}) is reducing to $\varrho(p)=\delta\left(1-\frac{1}{W}\right)$ (microcanonical). Under this $\varrho$, generalized entropies take the form of 
\begin{equation}
        S_g=W\left<g(p)\right>_{\varrho}=\sum^k_{j=1}V_j g\left(\frac{V_j}{\sum^k_{m=1}V^2_m}\right)
        \approx V^{*}g\left(\frac{1}{V^{*}}\right),
    \label{eq:asymp_multi}
\end{equation}
where $V^{*}=\max_j V_j$ defines the asymptotically leading term. In spite of the obvious analogy between Eq. (\ref{eq:micro_entropy}) and the last part of Eq. (\ref{eq:asymp_multi}), the two expressions can fairly differ from each other, e.g. in terms of non-extensivity classes. In the next paragraph, we proceed with providing a higher resolution picture of entropies defined under multi-delta density.

The entropy form in Eq. (\ref{eq:asymp_multi}) implicitly suggests that if a particular region $V_0$ of the configuration space is forbidden, or the regions  $V_{i=1,\dots,k}\neq V^{*}$ are rarely visited, then the generalized entropies are principally determined by the behaviour of the dominant term $V^{*}$, which has already been discussed in various different contexts such as in Refs.\cite{HT_gen_ent_appl,ruseckas_binary_spin}.  This additionally implies that extensivity is entirely encoded in the specific dependence of $V^{*}=V^{*}(W)$. For simplicity
hereinafter we assume that the volume of the entire configuration space is scaling with the volume of the leading sub-domain as $V^{*}(W)\sim W^{\xi}$, yielding
\begin{equation}
    S_g\approx W^{\xi}g\left(\frac{1}{W^{\xi}}\right),
    \label{eq:multidelta_entropy_scaled}
\end{equation}
where $0<\xi\leq1$ for obvious reasons (see Supplementary Information S2). For the linear case ($\xi=1$) regular sample space scaling is recovered \cite{HT_gen_ent_class}, however, sub-linear dependence can have very interesting consequences in terms of the H-T scaling given in Eq. (\ref{eq:Hannel_Turner_scaling_base_def}). 
Plugging $V^{*}(W)\sim W^{\xi}$ into Eq. (\ref{eq:gen_scaling}), i.e. keeping track of how entropies change under the rescaling of the configuration space volume we obtain
\begin{align}
    R_{\lambda}=\lim\limits_{W\to\infty}\frac{V^{*}(\lambda W)g_q\left(\frac{1}{V^{*}(\lambda W)}\right)}{V^{*}(W)g_q\left(\frac{1}{V^{*}(W)}\right)}=\lim\limits_{W\to\infty}\frac{1-(\lambda W)^{\xi(1-q)}}{1-W^{\xi(1-q)}}\sim \lambda^{1-c},
\end{align}
where $c=1-\xi+\xi q$. Compared to the microcanonical ensemble where Tsallis entropy, $S_q$ is extensive under $W(N)\sim N^{\frac{1}{1-q}}$, here extensivity is satisfied if $W(N)\sim N^{\frac{1}{1-\xi+\xi q}}$. For the sake of detailed comparison of the scaling exponents under different forms of $\varrho$ see Table\ref{tab:bgs_tab} and Table\ref{tab:tsallis_tab}.  

\subsubsection*{A single macroscopic state (Bose-Einstein)}

In this example we consider a system comprising of a single state with a macroscopic large probability, $p_1=1-\frac{1}{W-1}$, together with $W-1$ evenly distributed further states whose corresponding probabilites are given by $p_{i=2,\dots,W}=\frac{1}{(W-1)^2}$. Based on the analogy with a \emph{Bose-Einstein condensate}, we refer to this system as Bose-Einstein, and write the corresponding density of states as
\begin{align}
    \varrho(p)=\frac{1}{W}\delta\left(p-\left(1-\frac{1}{W-1}\right)\right)+\frac{W-1}{W}\delta\left(p-\frac{1}{(W-1)^2}\right).
\label{eq:BEC}
\end{align}
Using that, generalized entropies can analytically be expressed as \begin{equation}
    S_g=g\left(1-\frac{1}{W-1}\right)+(W-1)g\left ( \frac{1}{(W-1)^2} \right ).
    \label{eq:entropy_bec}
\end{equation}
Note that Eq. (\ref{eq:entropy_bec}) differs greatly from its microcanonical analogue Eq. (\ref{eq:micro_entropy}), revealing that even few configurations with macroscopic probabilities can significantly alter the behaviour that we would predict based on the microcanonical ensemble.

A system hallmarked by the density of states given above in Eq. (\ref{eq:BEC}) is displaying strong heterogeneity in its configurations as the variance of configuration probabilities decays linearly with the expected value, $\sigma^2\stackrel{W\to\infty}{\sim}\left<p\right>\sim W^{-1}$, contrary to the microcanonical picture, where $\sigma^2\sim0\sim W^{-\infty}$. Due to the presence of macroscopically large weights, entropic forms have non-zero contributions coming from $p\approx1$, therefore H-T scaling is given by
\begin{equation}
    R_{\lambda}=\lim\limits_{W\to \infty}\frac{g\left(1-\frac{1}{\lambda W}\right)+\lambda W g\left(\frac{1}{(\lambda W)^2}\right)}{g\left(1-\frac{1}{W}\right)+Wg\left(\frac{1}{W^2}\right)}\sim \lambda^{1-c},
\end{equation} 
where $c=2$ for BGS and $c=2q$ for Tsallis entropies (see Supplementary Information S2). Note that in both cases the $c$ exponent is greater than that of corresponding to the microcanonical picture. Hence, this example apparently reveals how H-T scaling is changing as $\varrho(p)$ broadens. As another consequence, under the previous density of states, Tsallis entropy with deformation parameter $q<\frac{1}{2}$ is extensive if $W(N)\sim N^{\frac{1}{1-2q}}$ however, for $q>\frac{1}{2}$ extensivity of $S_q$ surprisingly requires $W(N)$ to be monotonically decreasing (see Table\ref{tab:bgs_tab} and Table\ref{tab:tsallis_tab}).

\subsubsection*{Exponential density of states (exponential)}
Another characteristic type of decays can be formulated through an exponential density of states, which is simply expressed as
\begin{equation}
\varrho(p)=We^{-Wp}.
\end{equation}
Although this density function satisfies the normalizaton condition only in the asymptotic sense, i.e., $\int^1_{0}\varrho(p) \mathrm{d}p=1+\mathcal{O}(e^{-W})$, the corresponding approximation error decays rapidly with $W$. Hence, the expected value of the above exponential form can safely be approximated by $\left<p\right>_{\varrho}= \frac{1}{W}+\mathcal{O}(e^{-W})$,   
in accordance with the constraints imposed in Results section. The Boltzmann-Gibbs-Shannon entropy in this case takes the form of
\begin{equation}
    S_{\text{BGS}}=W\left<-p\ln p\right>_{\varrho}\stackrel{W\to\infty}{\approx}\ln W,
\end{equation}
while the Tsallis entropy with $q<1$ deformation parameter can be given as
\begin{align}
    S_q=W\left<\frac{p-p^q}{q-1}\right>_{\varrho}= 
\frac{1-e^{-W}(W+1)}{q-1}+W^{-(q-1)}\frac{\Gamma(q+1,W)-\Gamma(q+1)}{q-1}
\label{eq:tsallisexp}
    \approx \frac{1}{q-1} \left(1- \Gamma(q+1)W^{1-q}\right),
\end{align}
where the equality is obtained by neglecting asymptotically vanishing terms. Although the aforementioned density of states satisfies the necessary conditions only asymptotically, the closely related form of $\varrho(p)=\lim_{p\to0}\left(W-1\right)\left(1-p\right)^{W-2}\sim We^{-Wp}$ has the nice property of exactly fulfilling both normalization and expected value constraints. For example, the BGS entropy in this case can be written as $S_{\text{BGS}}=\psi(W+1)-\psi(2)\approx \ln W-\frac{1}{2W}-\psi(2)$, where $\psi(x)=\frac{\mathrm{d} \ln \Gamma(x)}{\mathrm{d} x}$ denotes the Digamma function.

The results above show that an exponentially decaying $\varrho(p)$ is always resulting in regular scaling. This practically implies that all the results obtained for the microcanonical ensemble, including i.e. the H-T scaling properties can automatically be extended to the exponential case without any modifications (see Table\ref{tab:bgs_tab} and Table\ref{tab:tsallis_tab}).

\subsubsection*{Log-gamma density of states (log-gamma, limiting log-normal)}

A density of states that limits log-normal when $W\to\infty$ yet is defined on the bounded support $[0,1]$ can be defined as the distribution of the product of independent uniform variables on $[0,1]$. We call this distribution log-gamma, since in log-transformed variables, it is a sum of exponentially distributed variables, i.e., a gamma distribution. As the number of terms in the sum approaches infinity, gamma limits normal; consequently, as the number of terms in the product approaches infinity, log-gamma limits log-normal (for details, see \cite{dettmann2009product}). Its tail on a log-log scale (limiting a quadratic function, interpolating between an exponential and a power law tail) is shown in Figure \ref{fig2}a. The log-gamma density of states is 
defined as \cite{dettmann2009product} 
\begin{equation}
\varrho(p)=\frac{\ln^{w}\left(\frac{1}{p}\right)}{\Gamma (w+1)},
\end{equation}
where $w=\log_2 \left(W\right)-1$. 
Under the previous form of $\varrho$, BGS entropy reads (see Supplementary Information S2)
\begin{equation}
    S_{\text{BGS}}=W\left<-p\ln p\right>_{\varrho} = W\frac{w+1}{2^{w+2}} \sim \ln W, 
\end{equation}
corresponding to a regular (logarithmic) scaling with $W$.
In contrast, the Tsallis entropy surprisingly shows anomalous scaling for $q\neq 1$ as
\begin{equation}
    S_{q}=\frac{1}{q-1}\left(1-W^{1-c}\right),
\end{equation}
where $c=\log_2 \left(1+q\right)\neq q$. 
On one hand this means that the microcanonical and log-gamma density of states are practically indistinguishable from the point of view of BGS entropy, while on the other hand, it also implies that H-T exponent $c=\log_2 \left(1+q\right)$ associated with Tsallis entropies and their deformation parameters $q$ do not coincide anymore. Consequently, under log-gamma density of states extensivity of $S_q$ can be obtained for the systems where $W(N)\sim N^\frac{1}{\log_2\left(\frac{2}{1+q}\right)}$. In Figure \ref{fig2}c we show how $S_{q}$ scales with $W$ as a function of the deformation parameter for both microcanonical and log-gamma density of states.

\subsubsection*{Power law density of states (power law)}

Power-law like decay offers a much 
larger heterogeneity over the configuration space compared to e.g., the exponential density of states, and therefore, might dramatically alter the dependence of entropies upon $W$. To illustrate this, let us first 
define the corresponding density of states as 
\begin{equation}
\varrho(p)=\frac{1}{W-1}p^{-1+\frac{1}{W-1}}.
\end{equation}
Surprisingly, this slow decay characterized by the strong inhomogeneity of probabilities over the configuration space yields to an anomalous scaling of both BGS and Tsallis entropy,
\begin{align}
    S_{\text{BGS}}=1-\frac{1}{W} \ \ \  \text{and} \ \ \ 
    S_q=\frac{1}{q-1}\left(1-\frac{W}{qW-q+1}\right).
    \label{eq:pow_bgs} 
\end{align}
Note that BGS and Tsallis entropies remain finite even in the limit of $W\to \infty$, consequently their extensivity can not be carried out anymore, disenabling \emph{thermodynamical} description of the corresponding systems. In general, we proclaim that the impossibility of extensivity suggests the lack of \emph{disorder} and uncertainty.

\subsubsection*{Beta density of states (beta)}

A particularly interesting example of first order statistics is the normalized Beta distribution, representing family of continuous probability density functions that are parametrized by two positive shape parameters $a$ and $b$, written as
\begin{equation}
    \varrho(p)=\frac{1}{B(a,b)}p^{a-1}(1-p)^{b-1},
\label{eq:beta}
\end{equation}
where $B(a,b)$ denotes the Beta function. In order to satisfy the previously discussed expected value constraint, the shape parameters should algebraically be related to the sample space volume as $\left<p\right>_{\varrho}=\frac{a}{a+b}=\frac{1}{W}$, therefore $b=a(W-1)$. According to previous relation, either $a$ or $b$ could be chosen arbitrarily while the correct choice for the other shape parameter has to be made in the light of the previous one. Under these settings BGS entropy can simply be expressed as
\begin{equation}
    S_{\text{BGS}}=W\left<-p\ln p\right>_{\varrho}=\psi(aW+1)-\psi(a+1),
    \label{eq:digamma}
\end{equation}
where $\psi(x)=\frac{\mathrm{d} \ln \Gamma(x)}{\mathrm{d} x}$ denotes the Digamma function. When $a$ is in the regime of $a\sim\frac{1}{W-1}\ll1$, the distribution becomes extremely skewed, resulting in anomalous scaling,
\begin{equation}
S_{\text{BGS}}\sim 1-\frac{1}{W} 
\end{equation}
whereas greater values $a\gg 1$ yield regular $S_{\text{BGS}}\sim \ln W$ scaling. 
The Tsallis entropy takes the form of 
\begin{equation}
    S_q=\frac{1}{q-1}\left(1-W\frac{B\left(aW,q\right)}{B(a,q)}\right),
\end{equation}
which again results in anomalous scaling for $a\sim\frac{1}{W-1}$. 

According to the above, first order statistics following the Beta distribution can lead to substantially different types of configuration space scaling depending on the specific choice for the shape parameter $a=a(W)$. This can be seen by letting $b\approx aW$ and then using the approximate form of Eq. (\ref{eq:beta}) which reads
\begin{equation}
    \varrho(p)\stackrel{p\to0}{\sim}p^{-1+a}e^{-Wap}
\label{eq:beta_approx}
\end{equation}
with the free parameter of $a(W)$ controlling the broadness of the distribution. By tuning the shape parameter in Eq. (\ref{eq:beta_approx}) above, we can smoothly interpolate between an exponential and power law like $\varrho(p)$, and in parallel keep track of how our system virtually undergoes a transition from a highly disordered state (hallmarked by exponential density) to the ordered regime (characterized by power-law density).

\begin{figure}[htp]
  \begin{center}
  \centering
    \includegraphics[width=1.0\textwidth]{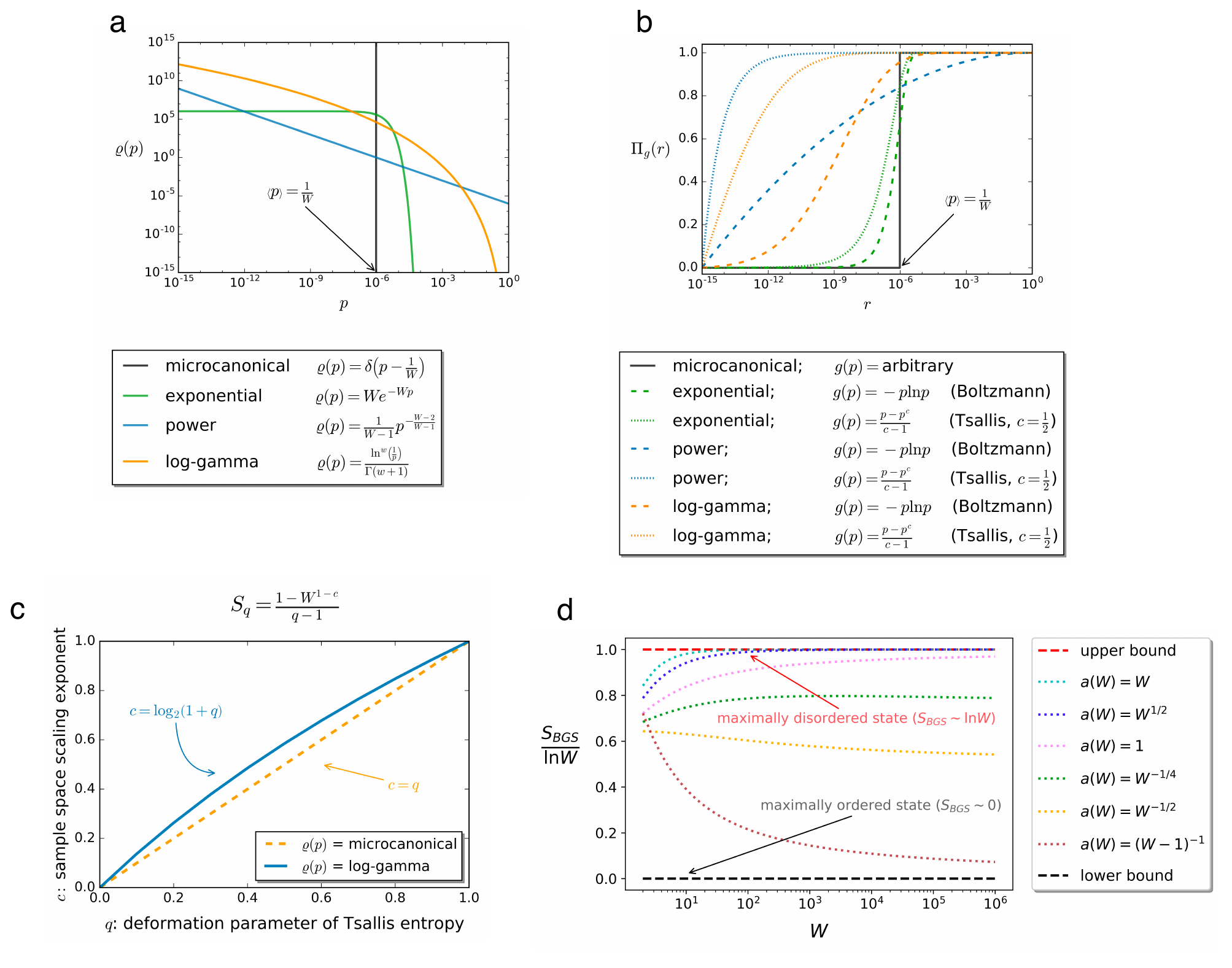}
      \caption{\textbf{a)} Continuous, parameter-free density of states we consider in this paper. \textbf{b)} Contribution of configurations with probability $p<r$ to the total entropy of the system, which we call cumulative entropy $\Pi_g(r)$, for different combinations of density of states $\varrho$ and entropy kernels $g$. 
      In each case, $\left<p\right>_{\varrho}=W^{-1}=2^{-20}\approx10^{-6}$. \textbf{c)} Hanel-Thurner exponent, given by Eq. (\ref{eq:gen_scaling}), of Tsallis entropies for the microcanonical ensemble and log-gamma (limiting log-normal) density of states. Although the H-T exponent of BGS entropy ($q=1$) is invariant to changing the density of states from microcanonical to log-gamma, this is no longer true for Tsallis entropies ($q<1$), indicating that non-extensivity class of any system is jointly determined by its extensive generalized entropy and the system's density of states. \textbf{d)} Scaling of BGS entropy $S_{\text{BGS}}$ with configuration space size $W$ when the system's the density of states follows a beta distribution with a power-law tail, characterized by $a(W)$. Note that in systems with $a\sim W^{-\epsilon}$ with $\epsilon \geq 1$, BGS entropy converges to a finite value in the thermodynamic limit $W\to\infty$.
    }
    \label{fig2}
    \end{center}
\end{figure}

We proceed with providing a detailed description of the above transition. For simplicity, here we restrict our analysis solely to the BGS entropy, however the results can straightforwardly be extended to the Tsallis entropy. 
If $a(W)>\frac{1}{W-1}$,
the first term in Eq. (\ref{eq:digamma}) is dominating the second term, and based on the asymptotic expansion of the Digamma function we recover the regular $S_{\text{BGS}}\approx \psi(aW+1)\sim \ln W$ scaling. If however, $a(W)<\frac{1}{W-1}$, the density of states takes a power law like form, and the system is governed into a highly ordered configuration. The very existence of this transition between the two different regimes is mathematically encoded in the properties of the Digamma function. At the transition line $a(W)=\frac{1}{W-1}$ between the two regimes 
\begin{align}
    S_{\text{BGS}}=\psi\left(2+\frac{1}{W-1}\right)-\psi\left(1+\frac{1}{W-1}\right)=1-\frac{1}{W}.
\end{align}
Below this line the BGS entropy can not satisfy extensivity. In slightly more intuitive terms, the distribution of the configuration probabilities becomes so extremely skewed in this regime that practically no uncertainty is appearing in the system description, which therefore, cannot be made extensive. In Figure \ref{fig2}d. we illustrate the scaling of the BGS entropy in multiple different cases.

\newcommand\xrowht[2][0]{\addstackgap[.5\dimexpr#2\relax]{\vphantom{#1}}}
\begin{table*}[ht!]
\resizebox{1.0\textwidth}{!}{\begin{minipage}{\textwidth}
        \caption{Scaling relations between BGS entropy $S_{\text{BGS}}\sim W\left<-p\ln p\right>_{\varrho}$, configuration space size $W$, and effective system size $N$, along with Hanel-Thurner exponent $c$, summarizing the results of the paper. Detailed calculations, following the outline shown in Figure \ref{fig1}c, are given in the Results section and in the Supplementary Information S2.}
\begin{tabular}{|c|c|c|c|}
\hline \xrowht{20pt}
$\varrho(p)$ & $S_g(W)$ & $c$ & $ W(N)$
  \\ \hline\hline \xrowht{20pt}
 microcanonical;\ \ \ $\delta\left(p-\frac{1}{W}\right) $ & $\ln W $ & $1$ & $e^N$
 \\ \hline \xrowht{20pt}
 multi-delta;\ \ \ Eq. (\ref{eq:multidelta}) & 
$\xi\ln W$
 & $1$ & $e^N$
\\ \hline \xrowht{20pt}
 Bose-Einstein;\ \ \ Eq. (\ref{eq:BEC})& $\frac{W-2}{W-1}\ln\left(\frac{W-1}{W-2}\right)+\frac{2}{W-1}\ln\left(W-1 \right ) $ & $2$ & $e^{-\mathcal{W}_{-1}\left(-\frac{N}{2}\right)}$
\\ \hline \xrowht{20pt}
exponential;\ \ \ $We^{-Wp}$ & $\ln W$ & $1$ & $e^N$
\\ \hline \xrowht{30pt}
log-gamma;\ \ \ $\frac{\ln^{w}\left(\frac{1}{p}\right)}{\Gamma (w+1)}$ & $\ln W$ & $1$ & $e^N$
\\ \hline \xrowht{20pt}
power;\ \ \ $\frac{1}{W-1}p^{-1+\frac{1}{W-1}}$  & $1-\frac{1}{W}$ & $1$ & never extensive
\\ \hline \xrowht{30pt}
beta;\ \ \ $\frac{p^{a-1}(1-p)^{a(W-1)-1}}{B\left(a,a\left(W-1\right)\right)}$ & $\psi(aW+1)-\psi(a+1)$  & $\left\{\begin{matrix}
\ 1, \ \text{if} \ a\gg \frac{1}{W}\\ 
\ 0, \ \text{if} \ a\leq  \frac{1}{W}
\end{matrix}\right.$  & $\left\{\begin{matrix}
\ e^N, \ \text{if} \ a\gg \frac{1}{W}\\ 
\ \text{not extensive}, \ \text{if} \ a\leq  \frac{1}{W}
\end{matrix}\right.$
\label{tab:bgs_tab}
\\ \hline
\end{tabular}
\end{minipage}}
\end{table*}

\begin{table*}[ht!]
\resizebox{1.0\textwidth}{!}{\begin{minipage}{1\textwidth}
        \caption{Scaling relations between Tsallis entropies $S_{q}\sim W\left<\frac{p-p^q}{q-1}\right>_{\varrho}$, configuration space size $W$, and effective system size $N$, along with Hanel-Thurner exponent $c$, summarizing the results of the paper. Detailed calculations, following the outline shown in Figure \ref{fig1}c, are given in the Results section and in the Supplementary Information S2.}
\begin{tabular}{|c|c|c|c|}
\hline \xrowht{20pt}
$\varrho(p)$ & $S_g(W)$ & $c$ & $ W(N)$
  \\ \hline\hline \xrowht{20pt}
 microcanonical;\ \ \ $\delta\left(p-\frac{1}{W}\right) $ & $\frac{1-W^{1-q}}{q-1}$ & $q$ & $N^{\frac{1}{1-q}}$
 \\ \hline \xrowht{30pt}
 multi-delta;\ \ \ Eq. (\ref{eq:multidelta}) & 
 $\frac{1-\left(W^{\xi}\right)^{1-q}}{q-1}$
 & $1-\xi+\xi q$ & $N^{\frac{1}{\xi(1-q)}}$
\\ \hline \xrowht{30pt}
 Bose-Einstein;\ \ \ Eq. (\ref{eq:BEC}) & $\frac{1-\left(1-\frac{1}{W-1} \right )^q-\left(\frac{1}{W-1} \right )^{2q-1}}{q-1}$ & $2q$ & $N^{\frac{1}{1-2q}}$
\\ \hline \xrowht{20pt}
 exponential;\ \ \ $We^{-Wp}$ & $\frac{1-W^{1-q}}{q-1}$ & $q$ & $N^{\frac{1}{1-q}}$
\\ \hline \xrowht{30pt}
 log-gamma;\ \ \ $\frac{\ln^{w}\left(\frac{1}{p}\right)}{\Gamma (w+1)}$ & $\frac{1-W^{\log_2\left(\frac{2}{1+q}\right)}}{q-1}$ & $\log_2\left(1+q\right)$ & $N^{\frac{1}{\log_2\left(\frac{2}{1+q}\right)}}$
\\ \hline \xrowht{20pt}
power;\ \ \ $\frac{1}{W-1}p^{-1+\frac{1}{W-1}}$  & $1-\frac{W}{qW-q+1}$ & $1$ & never extensive
\\ \hline \xrowht{30pt}
beta;\ \ \ $\frac{p^{a-1}(1-p)^{a(W-1)-1}}{B\left(a,a\left(W-1\right)\right)}$  &  $\frac{1}{q-1}\left(1-W\frac{B\left(aW,q\right)}{B(a,q)}\right)$ & $\left\{\begin{matrix}
\ q, \ \text{if} \ a\gg \frac{1}{W}\\ 
\ 0, \ \text{if} \ a\leq  \frac{1}{W}
\end{matrix}\right.$  & $\left\{\begin{matrix}
\ N^{\frac{1}{1-q}}, \ \text{if} \ a\gg \frac{1}{W}\\ 
\ \text{not extensive}, \ \text{if} \ a\leq  \frac{1}{W}
\end{matrix}\right.$
\label{tab:tsallis_tab}
\\ \hline
\end{tabular}
\end{minipage}}
\end{table*}

\section*{Discussion}
\label{sect:discussion}

Any dynamics, microscopic and coarse-grained, transient and stationary, takes place in the space of all configurations of a system. The size of the configuration space, $W$, is controlled by the system size $N$. For systems composed of weakly interacting variables, $W$ scales exponentially with $N$; the two provides a synonymous description of system size.
System size independent (statistical) models can be formulated in terms of homogeneous (e.g., intensive or extensive) functions of $N$, without paying much attention to the fact that dynamics actually takes place on configuration space. 

In complex systems of many strongly interacting variables, this is no longer true. The size of the configuration space might scale non-exponentially with system size, depending on multiple, not yet fully understood facets of correlated, history-dependent dynamics. A macroscopic attempt to extend ideas of statistical mechanics, in particular, the idea of size-invariance, formulated in terms of well-defined thermodynamic limits, to such complex systems can be built on the concept of generalized entropies $S_g$. Generalized entropies connect system size $N$ with configuration space size $W$ by implicitly re-defining system size as proportional to the generalized entropy characteristic to the system. Generalized entropies are, therefore, extensive by definition, i.e., $S_g\sim N$. Note that this inverse logic is necessary as the space of dynamics is the configuration space; system size, of which homogeneous functions and system-size invariant models can be formulated, is auxiliary. 

Generalized entropies account for higher-order statistics of the dynamics over configuration space, resulting in history-dependence (mathematically formulated as e.g., non-ergodicity), or correlated visiting probabilities. What they do not describe is visiting probabilites themselves, reflected by the fact that $S_g$ takes these visiting probabilities as arguments. Conventional statistical mechanics suggest that changing visiting probabilities, i.e., the distribution over configurations, does not alter the (asymptotic) relation between system size and configuration space size: the Boltzmann-Gibbs-Shannon entropy $S_{\text{BGS}}$ is extensive in all known ensembles. 

In this paper, we show that this is not the case. First order statistics of visiting probabilities, formulated in terms of density of states $\varrho$, might very well change the relation between $N$ and $W$. We identify three classes of density of states. Class $i)$ does not change the asymptotic scaling of $W(N)$ compared to the uniform (microcanonical) distribution over configuration space $\varrho=\delta(p-W^{-1})$. These density of states we call regular, as defined in Eq. (\ref{eq:regularanomalous}).

Class $ii)$ entails density of states that change $W(N)$ asymptotically (i.e., they are anomalous, according to Eq. \ref{eq:regularanomalous}) yet an extensive generalized entropy can be still assigned. This extensive generalized entropy is \emph{necessarily a different one} that is extensive for the microcanonical case. The appropriate generalized entropy is thus dependent on the visiting probabilities, e.g., on the thermodynamic ensemble in equilibrium statistical physics. This is observed in a $\varrho$ corresponding to a system with a microstate with macroscopic probability (that we call Bose-Einstein, BE) in both weakly and strongly correlated systems, modeled by $S_{\text{BGS}}$ and by the family of Tsallis entropies, $S_q$, respectively. First order statistics also modify the non-extensivity scaling $W(N)$ in strongly correlated systems in case of a $\varrho$ limiting log-normal, and a $\varrho$ is corresponding to multiple uniform domains in configuration space, whereas they do not modify $W(N)$ in weakly correlated systems.

The third class describes density of states that are so fat-tailed that $W(N)$ saturates, and therefore, no extensive entropy can be assigned to such systems. We find that systems with a power-law density of states, with exponent approaching $-1$ in the limit of $W\to\infty$ belong to this class, regardless of their higher-order statistics. Such systems have non-zero visiting probability at all configurations, yet these visiting probabilities are so skewed that both $S_{\text{BGS}}$ and $S_q$ saturates at a finite value when $W\to\infty$. 

We believe that this unified macroscopic picture of strongly interacting and history-dependent processes, based on first and higher order statistics of visiting probabilities over configuration space, makes statistical mechanics a more flexible tool for modeling complex systems both within and outside the realm of physics.

\renewcommand{\thefigure}{S\arabic{figure}}
\renewcommand{\thetable}{S\arabic{table}}
\renewcommand{\theequation}{S\arabic{equation}}
\renewcommand{\thesection}{S\arabic{section}}
\setcounter{equation}{0}
\renewcommand{\thesubsection}{S\arabic{subsection}}
\renewcommand{\thesubsubsection}{s\arabic{subsubsection}}

\section*{Supplementary Information}
\subsection{Generalized entropies and Jensen inequality}
Generalized entropies defined by Eq. (\ref{eq:rho_gen_entropies}) are strictly positive $S_g\geq0 \ \forall  \{ g(p),\varrho(p)\}$ since $g(p),\varrho(p)\geq0 \ \forall p\in[0,1]$. Besides that, they satisfy the second Shannon-Khinchin axiom (SK2), namely that systems given in their most disordered states are characterized by equiprobable-distributions (entropy is maximal for the uniform distribution). By imposing concavity upon $g(p)$, this can easily be verified through the well-known Jensen's inequality (J.I.) for concave functions:
\begin{equation}
    S_g=W\left< g\left(p\right)\right >_{\varrho} \stackrel{\text{J.I.}}{\leq}W g\left(\left< p \right>_{\varrho}\right)
    =Wg\left(\frac{1}{W}\right)=W\left< g(p)\right >_{\varrho=\delta(p-1/W)}=\max_{\varrho}S_g,
\end{equation}
where we have exploited 
both normalization and expected value constraints.

\subsection{Density of states: examples}

\subsubsection{Uniform distribution over configuration space (microcanonical)}

Despite its triviality the microcanonical ensemble is unambiguously among the most relevant examples since it is an essential part of thermodynamics, statistical physics as well as information theory. 

It is defined by uniform probabilities over the configuration space, that is $\varrho=\delta(p-1/W)$, whose normalization can simply be deduced from the definition of Dirac delta function. Correspondingly, expected value condition is given by
\begin{equation}
    \left<p\right>_{\varrho}=\int\limits^1_{0}p\cdot \delta\left(p-\frac{1}{W}\right) \mathrm{d}p=\frac{1}{W}
\end{equation}
in agreement with Eq. (\ref{eq:rho_exp_val_const}).
 
\subsubsection{Multiple uniform domains over configuration space (multi-delta)}

Multi-delta is the generalization of the microcanonical picture, where different sets of configurations are given. Within these sets each configuration are equiprobable however, configuration belonging to different sets might have different probabilites assigned to them.

The corresponding $\varrho(p)$ defined by Eq. (\ref{eq:multidelta}) is normalized since
\begin{equation}
\int \varrho(p) \mathrm{d}p=0+\sum^k_{j=1}\frac{V_j}{W}=1,
    \end{equation}
whereas expected value constraint can be verified by evaluating
\begin{equation}
\left<p\right>_{\varrho}= \int p\varrho(p) \mathrm{d}p=\frac{V_0}{W}\cdot 0+\sum^k_{j=1}\frac{V_j}{W}\frac{V_j}{\sum\limits^k_{m=1}V^2_m}=\frac{1}{W}.
\end{equation}
Once normalization and expected value conditions are imposed, one can easily check that under the multiplicative rescaling of confiuration space volume BGS entropy scales as
\begin{align}
    R_{\lambda}=\lim\limits_{W\to\infty}\frac{V^{*}(\lambda W)g_\text{BGS}\left(\frac{1}{V^{*}(\lambda W)}\right)}{V^{*}(W)g_\text{BGS}\left(\frac{1}{V^{*}(W)}\right)}=\lim\limits_{W\to\infty}\lambda^{\xi}\frac{\frac{1}{(\lambda W)^{\xi}}\ln \frac{1}{(\lambda W)^{\xi}}}{\frac{1}{W^{\xi}}\ln \frac{1}{W^{\xi}}}\sim \text{const.},
\end{align}
implying that $c=1$ in accordance with the corresponding row in Table\ref{tab:bgs_tab}. Consequently, multi-delta density of states can not change the $c$ exponent, therefore, BGS entropy remains extensive under $W(N)\sim e^N$. 

Considering the case of Tsallis entropy, we take the $p\to0^{+}$ limit (for $\xi>0$) and utilize the expansion of $g_q(p)$ around $p=0$
\begin{equation}
    g_q(p)=\left\{\begin{matrix}
\ \ \ \ \ \frac{1}{1-q}p^q+\mathcal{O}(p) \ \ \ \ \ \text{if} \ \ 0<q<1
\\
\\ 
-\frac{1}{1-q}p+\mathcal{O}(p) \ \ \ \ \text{if} \ \ q>1,
\end{matrix}\right.
\label{eq:jose_maria_tsallis}
\end{equation}
based on which we obtain
\begin{equation}
    R_{\lambda}\approx\frac{V^{*}(\lambda W)\left(\frac{1}{V^{*}(\lambda W)}\right)^q}{V^{*}(W)\left(\frac{1}{V^{*}(W)}\right)^q}
\end{equation}
for $0<q<1$. Assuming the scaling form of $V^{*}(W)\sim W^{\xi}$ the previous equation can be further simplified to
\begin{equation}
    R_{\lambda}=\frac{(\lambda W)^\xi\left(\frac{1}{(\lambda W)^\xi}\right)^q}{W^\xi\left(\frac{1}{W^{\xi}}\right)^q}\sim \lambda^{1-c},
\end{equation}
where $c=1-\xi+\xi q$ in agreement  with the corresponding row in Table\ref{tab:tsallis_tab}.

\subsubsection{A single macroscopic state (Bose-Einstein)}

Owing to the presence of a macroscopic state, H-T scaling and thus, conditions of extensivity substantially change under Bose-Einstein density of states. This stems from the fact that states with probabilites $p\approx 1$ have non-negligible contribution to the entropy compared to, e.g. the microcanonical picture where the major contribution is coming from $p=\frac{1}{W}\approx0$.

The corresponding density of states written in the form of Eq. (\ref{eq:BEC}) fulfills both normalization and expected value contraints since the following identities hold
\begin{equation}
    \int \varrho(p) \mathrm{d}p=\frac{1}{W}+\frac{W-1}{W}=1 \ \ \ \ \text{and} \ \ \ \
    \left<p\right>_{\varrho}=\frac{1}{W}\left(1-\frac{1}{W-1}\right)+\frac{W-1}{W}\frac{1}{(W-1)^2}=\frac{1}{W}.
\end{equation}
Plugging Eq. (\ref{eq:BEC}) into Eq. (\ref{eq:rho_gen_entropies}) and exploiting the fact that
\begin{equation}
    \int f(x)\delta(x-x_0)\mathrm{d}x=f(x_0)
\end{equation}
with $\delta(x)$ being the Dirac-delta function, we simply arrive to
\begin{equation}
    S_g=g\left(1-\frac{1}{W-1}\right)+(W-1)g\left ( \frac{1}{(W-1)^2} \right ).
\label{eq:bec_rho_supporting}
\end{equation}
Specifically, for the BGS entropy we obtain
\begin{equation}
    S_{\text{BGS}}=\frac{W-2}{W-1}\ln\left(\frac{W-1}{W-2}\right)+\frac{2}{W-1}\ln\left(W-1 \right ). 
\end{equation}
With the notations of $x=\frac{1}{W-1}$ and $z=\frac{1}{\lambda}$ the generalized $c$ exponent for BGS entropy is given by directly plugging Eq. (\ref{eq:bec_rho_supporting}) into Eq. (\ref{eq:gen_scaling})
yielding
\begin{align}
   R_{z}=&\lim\limits_{x\to 0^+}\frac{g_\text{BGS}\left(1-zx\right)+\frac{1}{zx}g_\text{BGS}\left((zx)^2\right)}{g_\text{BGS}\left(1-x\right)+\frac{1}{x}g_\text{BGS}\left(x^2\right)}
   =\lim\limits_{x\to 0^+}\frac{\left(1-zx\right)\ln(1-zx)+zx\ln\left((zx)^2\right)}{\left(1-x\right)\ln(1-x)+x\ln\left(x^2\right)}\nonumber \\ 
   \approx &\frac{z-2z\ln (zx)}{1-2\ln (x)}\sim z \equiv z^{c-1},
\end{align}
where $c=2$. Interestingly, extensivity of BGS can be maintained by imposing $W(N)\sim e^{-\mathcal{W}_{-1}\left(-\frac{N}{2}\right)}$. The previous form of $W(N)$ is a decreasing function of $N$ implying that asymptotically, $N\to\infty$, the configuration space has to collapse in order for extensivity to be satisfied. 

Under Bose-Einstein density of states Tsallis entropy reads
\begin{equation}
    S_q=\frac{1-\left(1-\frac{1}{W-1} \right )^q-\left(\frac{1}{W-1} \right )^{2q-1}}{q-1}.
    \label{eq:bec_tsallis}
\end{equation}
Based on Eq. (\ref{eq:bec_tsallis}) and Eq. (\ref{eq:jose_maria_tsallis}) generealized H-T scaling takes the form of
\begin{equation}
    R_z\approx\lim_{x\to 0^{+}}\frac{zx-(zx)^{2q-1}}{x-x^{2q-1}}\sim z^{2q-1}\equiv z^{c-1},
\end{equation}
where $c=2q$. For $q=\frac{1}{2}$, $S_q\to1$ therefore Tsallis entropy can not be made extensive. In case of $q\in \left(0,\frac{1}{2}\right)$ extensivity is obtained by setting $W(N)\sim N^{\frac{1}{1-2q}}$. If however, $q>\frac{1}{2}$ $S_q\sim N$ requires $W(N)$ to decrease with $N$.   

\subsubsection{Exponential density of states (exponential)}

Under exponential form of $\varrho$ noramlization condition gives
\begin{equation}
    \int\limits^1_0 We^{-Wp} \mathrm{d}p=1-e^{-W}\approx 1,
\end{equation}
while its expected value is simply written as
\begin{equation}
    \left<p\right>_{\varrho}= \int\limits^1_0 pWe^{-Wp} \mathrm{d}p=\frac{1-e^{-W}(W+1)}{W}\approx \frac{1}{W},
\end{equation}
where the approximations are obtained by neglecting asymptotically vanishing terms. Under this form of $\varrho$ BGS entropy is given by
\begin{equation}
    S_{\text{BGS}}=\ln W +\text{Shi}(W)-\text{Chi}(W)+e^{-W}-1+\gamma  
    \label{eq:sup_exp_bgs}
\end{equation}
where $\gamma$ is the Euler-Mascheroni constant while  Shi$(x)$ and Chi$(x)$ denote the hyperbolic sine and cosine integrals respectively. In the asymptotic limit of $W\to \infty$, Eq. (\ref{eq:sup_exp_bgs}) reduces to the regular $S_{\text{BGS}}\approx \ln W$ form. 

Moreover, Tsallis entropy with deformation parameter $q$ takes the form of
\begin{equation}
    S_{q}=W\left<\frac{p-p^q}{q-1}\right >_{\varrho}
=W^{-(q-1)}\frac{\Gamma(q+1,W)-\Gamma(q+1)}{q-1}+\frac{1-e^{-W}(W+1)}{q-1},
\label{eq:tsallisexp_alt}
\end{equation}
with $\Gamma(x,y)$ and $\Gamma(x)$ denoting the Incomplete gamma and Gamma functions respectively.
The previous equality along with the asymptotic expressions of $\Gamma(q+1,W\to\infty)=0$ and $\lim\limits_{W\to\infty}\frac{e^{-W}(W+1)}{q-1}=0$ suggest
\begin{equation}
    S_{q<1}\stackrel{W\to\infty}{=}  \frac{1-\Gamma(q+1)W^{1-q}}{q-1}.
    \label{eq:supptsallisexpasymptotic}
\end{equation}
Exponential distribution represents a particularly fast form of decay implying that the major contribution to entropies is coming from configurations where 
$p\approx W^{-1}$. This is supported by the fact that the corresponding green curves in Figure \ref{fig2}b are quite close to the Heaviside step function (realizing the microcanonical ensemble). As a consequence, requirements for extensivity do not change and the H-T exponents given by Eq. (\ref{eq:gen_scaling}) happen to have the same values as in case of microcanonical ensemble (for details see Table\ref{tab:bgs_tab} and Table\ref{tab:tsallis_tab}). 

\subsubsection{Log-gamma density of states (log-gamma, limiting log-normal)}

The appropriate parametrization of the log-gamma density of states is being carried out by setting up the following definite integral
\begin{equation}
    I_{\text{Log}}(m,w)=\int\limits^1_{0}p^m\frac{\ln^w\left(\frac{1}{p} \right )}{\Gamma(w+1)}\mathrm{d}p=\left(m+1\right)^{-(w+1)},
    \label{eq:sup_log_int}
\end{equation}
based on which the noramlization condition can automatically be verified by plugging $m=0$ into Eq. (\ref{eq:sup_log_int}). Note that if however, $m=1$ we obtain $\left<p\right>=2^{-w-1}$ which results in fulfilling expected value constraint only in case of $w=\log_2(W)-1$. 

Boltzmann-Gibbs-Shannon entropy is simply related to the evaluation of a similar integral, namely
\begin{equation}
    S_{\text{BGS}}= W\int\limits^1_{0}p\frac{\ln^{w+1}\left(\frac{1}{p}\right)}{\Gamma(w+1)} \mathrm{d}p=W(w+1)\int\limits^1_{0}p\frac{\ln^{w+1}\left(\frac{1}{p}\right)}{\Gamma(w+2)} \mathrm{d}p=\frac{W(w+1)}{2^{w+2}}.
\end{equation}
Using the fact that $w=\log_2(W)-1$ Shannon entropy given in its simplest form
\begin{equation}
    S_{\text{BGS}}=W\frac{\log_2(W)}{2W}\sim \ln W.
\end{equation}

Note that expected value constraint generally requires the precise evaluation of $I_{\text{Log}}(1,w)$ while Tsallis entropy can be expressed as
\begin{align}
    S_q=W\frac{I_{\text{Log}}(1,w)-I_{\text{Log}}(q,w)}{q-1}=\frac{1-W(q+1)^{-\log_2(W)}}{q-1},
\end{align}
which altogether with the identity of $1+q=2^{\log_2(1+q)}$ yield to
\begin{equation}
    S_q=\frac{1-W^{1-\log_2(1+q)}}{q-1}\equiv \frac{1-W^{1-c}}{q-1},
\end{equation}
where $c=\log_2(1+q)$ with the corresponding curve depicted in Figure \ref{fig2}c. As a surprising consequence, extensivity of the Tsallis entropy requires $W(N)\sim N^\frac{1}{\log_2(1+q)}$.

\subsubsection{Power-law density of states (power-law)}

For the sake of technical controllability, let us define the following integral
\begin{equation}
    I_{\text{Pow}}(A,m)=A\int\limits^1_{0} p^m \mathrm{d}p=\frac{A}{m+1},
    \label{eq:sup_pow_int}
\end{equation}
based on which we can precisely evaluate various expressions later on. First, with the substitutions of $A=\frac{1}{W-1}$ and $m=-1+\frac{1}{W-1}$ we obtain the normalization condition, whereas expected value requirement can be checked through $I_{\text{Pow}}\left(\frac{1}{W-1},\frac{1}{W-1}\right)=\frac{(W-1)^{-1}}{(W-1)^{-1}+1}=\frac{1}{W}$. 

Tsallis entropy can also be expressed as a function of the integral appearing in Eq. (\ref{eq:sup_pow_int}), namely
\begin{equation}
    S_q=W\frac{I_{\text{Pow}}\left(\frac{1}{W-1},\frac{1}{W-1}\right)-I_{\text{Pow}}\left(\frac{1}{W-1},q-1+\frac{1}{W-1}\right)}{q-1}=\frac{1}{q-1}\left(1-\frac{W}{qW-q+1}\right),
\end{equation}
which reduces to BGS entropy if $q\to1$, therefore 
\begin{equation}
    S_\text{BGS}=\lim\limits_{q\to1}\frac{1}{q-1}\left(1-\frac{W}{qW-q+1}\right)=1-\frac{1}{W},
\end{equation}
in a perfect accordance with Eq. (\ref{eq:pow_bgs}).

An alternative formulation of power-law density of states is provided by the form of
\begin{equation}
    \varrho(p)=\frac{2}{W-4}\left(p^{-1+\frac{2}{W-2}}-1\right),
\end{equation}
which has the nice property of $\varrho(p=1)=0$ but offerring the same type of scaling with $W$,
\begin{equation}
    S_{\text{BGS}}=3-\frac{4}{W}.
\end{equation}

The fact that under power-law form of $\varrho$ both BGS and Tsallis entropies asymptotically converge to a finite value implies that these entropic forms can not display extensivity, not even in the limit of $W\to\infty$, therefore making
it impossible to provide a thermodynamical description of the corresponding system. 
\subsubsection{Beta density of states (beta)}

The Shannon entropy under Beta distribution of the configuration probabilities is given by the following integral
\begin{align}
        S_{\text{BGS}}=W\left<-p\ln p\right>_{\varrho}=W\int\limits^1_{0}\ln \left(\frac{1}{p}\right)\frac{ p^{a}(1-p)^{b-1}}{B(a,b)}\mathrm{d}p = W\frac{a}{a+b}\left[ \psi(a+b+1)-\psi(a+1) \right ],
    \label{eq:sup_betadigamma}
\end{align}
where $\psi(x)=\frac{\mathrm{d} \ln \Gamma(x)}{\mathrm{d} x}$ denotes the digamma function. By imposing expected value constraint we obtain $\left<p\right>_{\varrho}=\frac{1}{W}=\frac{a}{a+b}$ based on which Eq. (\ref{eq:sup_betadigamma}) can be further simplified to
\begin{equation}
    S_{\text{BGS}}=\psi(aW+1)-\psi(a+1).
\end{equation}
In addition to this, assuming scaling form for $a(W)\sim W^{\eta-1}$, we arrive to
\begin{equation}
    S_{\text{BGS}}=\psi(W^{\eta}+1)-\psi(W^{\eta-1}+1).
    \label{eq:sup_phase_trans}
\end{equation}
Note that if $\eta > 0$ then $S_{\text{BGS}}\approx \psi(W^\eta)-\psi(1) \sim \eta\ln W $ asymptotically. As $\eta \to 0^{+}$ and $a^{*}\to \frac{1}{W-1}\sim W^{-1}$ a phase transition occurs whose presence can directly be derived from the recurrence relation of the Digamma function $\psi(x+1)=\psi(x)+\frac{1}{x}$. More precisely at the critical point $a^{*}$, where $\varrho$ decays as a power-law and extensivity is no longer accessible, Eq. (\ref{eq:sup_phase_trans}) displays the anomalous $S_{\text{BGS}}\sim 1-\frac{1}{W}$ form. Above this transition point, beta distribution however, reduces to an exponential form under which anomalous scaling of the BGS entropy is obtained.   

If first order statistics follows a beta distribution Tsallis entropy is given
\begin{equation}
    S_q=\frac{1}{q-1}\left(1-W\frac{\Gamma(a+q)\Gamma(aW)}{\Gamma(a)\Gamma(aW+q)}\right)=\frac{1}{q-1}\left(1-W\frac{B\left(aW,q\right)}{B(a,q)}\right),
    \label{eq:tsallisbeta1}
\end{equation}
where we have exploited the definiton of the beta function. Further algebraic manipulations for $a=1$ where $\varrho(p)\sim \left(1-p\right)^W$ and $B(W,q)\stackrel{W\to\infty}{\sim} \Gamma(q)W^{-q}$ yield to an asymptotic approximation of Tsallis entropies
\begin{equation}
    S_q=\frac{1}{q-1}\left(1-W(W+q)\Gamma(1+q)W^{-(1+q)}\right)
\approx  \frac{1}{q-1}\left(1-\Gamma(1+q)W^{1-q}\right),
    \label{eq:tsallisbeta3}
\end{equation}
therefore giving H-T exponent $c=1-q$. For further details see Table\ref{tab:bgs_tab} and Table\ref{tab:tsallis_tab}.

\bibliography{Entropy_refs_corr}
\bibliographystyle{ieeetr}

\end{document}